\documentclass[a4paper]{jpconf}
\usepackage{graphicx}
\begin{document}
\title{Temporal Variations of High-Degree Solar p-Modes from GONG and MDI}

\author{Olga Burtseva$^1$, Sushant Tripathy$^1$, Rachel Howe$^1$, Kiran Jain$^1$, Frank
Hill$^1$, Richard Bogart$^2$ and Maria Cristina Rabello-Soares$^2$}

\address{$^1$ National Solar Observatory, Tucson, AZ, USA}
\address{$^2$ HEPL, Stanford University, CA, USA}

\ead{burtseva@noao.edu}

\begin{abstract}
We study temporal variations in the amplitudes and widths of high-degree acoustic  modes
in the quiet and active Sun by applying ring-diagram technique to the GONG+ and MDI
Dopplergrams during the declining phase of cycle 23. The increase in amplitudes and
decrease in line-widths in the declining phase of the solar activity is in agreement
with previous studies. A similar solar cycle trend in the mode parameters is also seen
in the quiet-Sun regions but with a reduced magnitude. Moreover, the amplitudes obtained
from GONG+ data show long-term variations on top of the solar cycle trend.        
\end{abstract}

\section{Introduction}
The global and local analysis of solar acoustic modes has shown that the mode 
amplitudes and lifetimes are anti-correlated with the solar activity level, and depend
strongly on the local magnetic flux \cite{Chaplin00,Rajaguru01,Howe04}. Most of  the
analyses have been done for average quantities without discrimination between quiet and
active areas, while the solar cycle behavior of the quiet-Sun regions is not well
understood. High-degree \textit{p}-mode lifetimes measured in the quiet-Sun regions
using cross-correlation analysis show variations with the activity cycle
\cite{Burtseva09a}. In the quiet Sun, no large-scale magnetic field concentrations, one
of the potential damping sources, are present on the surface. Another explanation could
be the activity-related variation of the convective properties near the solar surface.
The studies of the solar cycle variations of the size of the solar granules so far
arrive at contradictory conclusions \cite{Berrilli99,Saldana04}. An attempt to mask
strong surface activity and analyze high-degree \textit{p}-mode amplitudes in the
quiet-Sun regions at solar minimum and maximum indicated that the amplitude at solar
minimum is higher than that at solar maximum \cite{Burtseva09b}, however, the effect
introduced by the mask needs to be better understood. In this work we apply
ring-diagram analysis to eight years of data to characterize the high-degree acoustic
mode amplitudes and widths in the quiet and active Sun during declining phase of the
activity cycle.   

\section{Data analysis} 
The mode parameters of the solar acoustic oscillations analyzed in this work are
obtained from Global Oscillation Network Group (GONG) and Michelson Doppler Imager
(MDI) high-resolution Dopplergrams for the period from 2001 to 2009 using the standard
ring-diagram technique \cite{Hill88} of the GONG and MDI pipelines. The size of a
standard patch in the ring analysis is $15^o \times 15^o$. The most distant  patches
are centered at 52$^o$.5 from the disk center. More recently the pipeline developed for
the Helioseismic and Magnetic Imager (HMI) \cite{Bogart11}, applying both asymmetric
\cite{Basu99} and symmetric \cite{Haber00} profiles fitting to derive the mode
parameters, has also been used to analyze the MDI data, in the traditional MDI
patch set. The results from both procedures are consistent, thus, here we show the
symmetric profile fitting results only.

The Magnetic Activity Index (MAI) values are computed from MDI 96-minute magnetograms 
for the same time frame of Dopplergrams. The MAI is computed by averaging  unsigned flux
above 50 G and below 500 G to avoid contamination from noise in  the measurements of the
quiet-Sun flux values and saturation issues of the MDI  magnetograms \cite{Basu04}. We
define the quiet-Sun regions as those with MAI values below 5 G. 

As discussed in \cite{Howe04}, the amplitudes and widths fitted from the ring analysis
need to be corrected for center-to-limb, residual B$_0$-angle variations, and
duty-cycle dependences. In our analysis, GONG data were restricted to duty cycle of
70$\%$ and higher. In contrast, MDI data have high duty cycle and therefore no
selection criteria were used to reject the fitted parameteres. In this work, only disk
center patches were analyzed to avoid the foreshortening effects.

\section{Results and discussion}  
The results for amplitudes and widths shown in this paper are for a multiplet $\ell$ =
440, $n$ = 2 ($\nu$ = 3.2 mHz). We find that the correlation of this multiplet with
other multiplets in the 2.5$-$3.5 mHz frequency range over solar activity cycle is higher
than 70$\%$. 

\subsection{Activity-related trend}
The variations of amplitude and width as a function of time and MAI, computed from all 
patches and the quiet-Sun patches at the disk center using GONG data, are shown in
Figure 1. According to the Mt.Wilson sunspot index data (see Figure 2), during most of
the year 2008 the Sun was quiet with no or very few small sunspots appearing on the
solar surface. Therefore, we took average of year 2008 consisting of 13 Carrington
Rotations as a reference, and plotted the mode parameters relative to this mean value.
The amplitudes increase by $\sim 10\%$ from the high solar activity period in 2001 till
2004. After this  period, we find a long-term variation, but no clear association with
activity cycle. It is interesting to note that the mode amplitude, derived from
Variability of Solar Irradiance and Gravity Oscillations (VIRGO) and GONG data using
global helioseismology technique, shows similar wiggles though they are small in
amplitude relative to the extent of the solar cycle trend \cite{Salabert11}. If we
ignore the unexplained long-term variations, we can conclude that the mode amplitudes
obtained in our study increased from 2001 to 2008 by $\sim 22\%$ in all patches and by
$\sim 16\%$ in the quiet-Sun regions. The widths decreased from 2001 to about 2008 by
$\sim 9\%$, then started showing a rising trend. 

\begin{figure}[t]
\begin{center}
\includegraphics[width=0.95\linewidth]{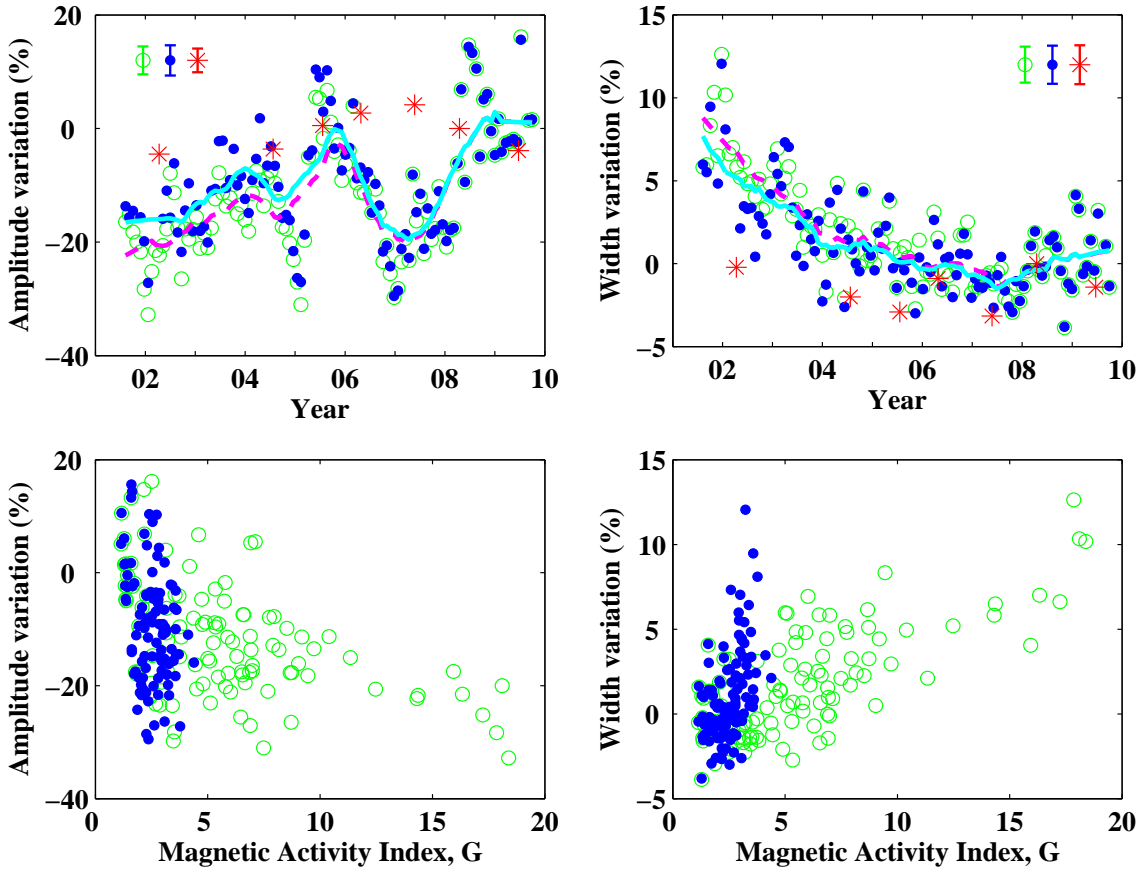}
\end{center}

\caption{Variation of amplitude and width (relative to the mean value of 2008) as a
function of time and MAI, computed in the quiet-Sun (closed blue circles) and all
patches without MAI threshold (open green cicrcles). The mode parameters are derived
from disk center patches using GONG data. Each point is an average over one Carrington
rotation. Solid light-blue and dashed magenta curves represent the temporally smoothed
values in the quiet-Sun and all patches, respectively. Red stars show the amplitude and
width variation as a function of time, computed from  MDI data in all patches at the
disk center. In this case, each point is an average over one year. The typical error
bars for the measurements are shown in the top panels.}

\end{figure}

\begin{figure}[t]
\begin{center}
\includegraphics[width=0.95\linewidth]{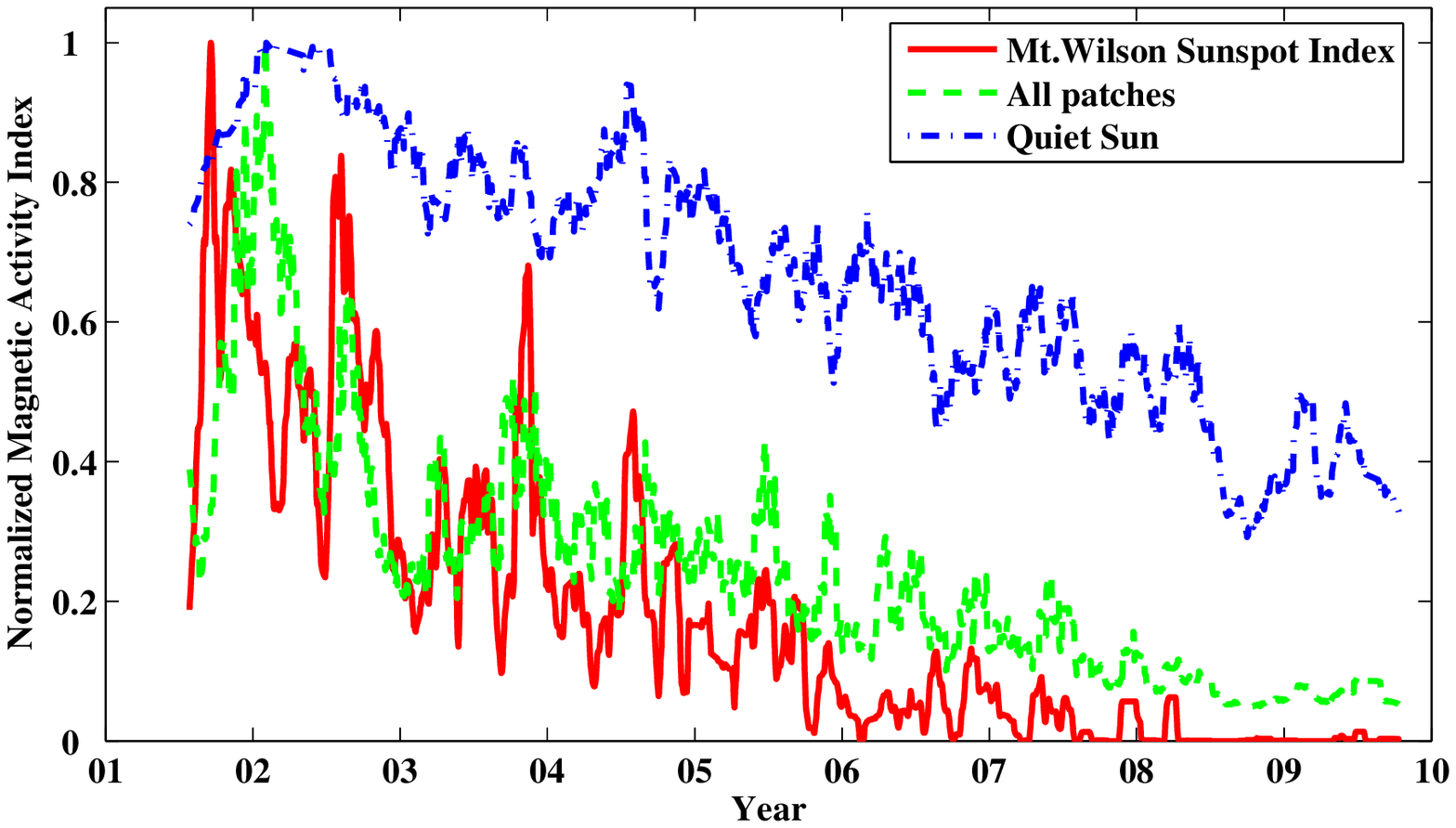}
\end{center}
\caption{Normalized magnetic activity index, computed from MDI magnetograms in the
quiet-Sun (dash-dot blue curve) and all regions (dashed green curve) at the disk
center, plotted as a function of time. The solid red line represents  the Mt.Wilson
Sunspot Index. Note, that each curve is normalized to its maximum value to show the
relative variation.}
\end{figure}

The variations in amplitude and width from MDI data (see Figure 1, top panels) show a
solar cycle trend similar to that from GONG data. The long-term  variations in the
amplitudes obtained from GONG data are not seen in the MDI  amplitudes. However, this
could be due to the limited amount of data (only 135 days) that were analyzed
with the MDI Dopplergrams. We plan to improve statistics on MDI data to confirm these
results.

The decrease in amplitudes and increase in widths with increase in MAI (bottom panels
of Figure 1) are consistent with known results from global and local analysis. The
linear correlation analysis shows that the correlation coefficient between amplitude
and MAI is $-0.49$ for all patches and $-0.29$ for the quiet-Sun patches. The
correlation coefficient between width and MAI is 0.73 for all patches and 0.48 for the
quiet-Sun patches.  

Figure 2 shows the variation of the magnetic activity index over time. The MAI values
of all patches at the disk center are well correlated with Mt. Wilson sunspot index.
A weaker solar-cycle trend is also visible in the magnetic indices of the quiet-Sun
patches. The solar cycle variation of the MAIs of the quiet-Sun regions computed from
MDI magnetograms was also noted in \cite{Burtseva09a} and could be due solar cycle
variation of the strong-field component of the quiet-Sun network
\cite{Hagenaar03,Pauluhn03}. This suggests that the magnetic field plays a role in
the activity related variations of the acoustic mode parameters in the quiet Sun. 

\subsection{Long-term variations}
In an attempt to understand the long-term variations, on top of the solar cycle trend,
seen in the amplitudes computed from GONG data, we restrict our analysis to 2 G $\le$
MAI $\le$ 3 G patches. We notice the long-term variation in the amplitude but no
dependence on magnetic activity. This leads to the conclusion that the variations are
probably not related to solar activity cycle. Moreover, according to our preliminary
estimates from the patches at higher latitudes, the variations are present up to the
regions centered at 52$^o$.5 in latitude. We plan to investigate this aspect in more
detail after applying the necessary geometrical corrections.

\ack
This work utilizes data obtained by the Global Oscillation Network Group (GONG)
Program, managed by the National Solar Observatory, which is operated by AURA, Inc.
under a cooperative agreement with the National Science Foundation. The data were
acquired by instruments operated by the Big Bear Solar Observatory, High Altitude
Observatory, Learmonth Solar Observatory, Udaipur Solar Observatory, Instituto de
Astrofisica de Canarias, and Cerro Tololo Interamerican Observatory. SOHO is a mission
of international cooperation  between ESA and NASA.

\end{document}